\documentclass[12pt]{article}
\usepackage[all]{xy}
\usepackage{array}
\usepackage{ifthen}
\usepackage{amscd}
\usepackage{amsmath}
\usepackage{amsfonts}
\usepackage{amssymb}
\usepackage{latexsym}
\usepackage{theorem}
\usepackage{pgf}
\usepackage{cancel}

\definecolor{cblack}{rgb}{0,0,0}
\definecolor{cverde}{rgb}{0,.5,0}
\definecolor{light}{rgb}{.8,.8,.8}
\definecolor{dark}{rgb}{.5,.5,.5}

\newcommand{\za}{\alpha}
\newcommand{\zb}{\beta}
\newcommand{\zd}{\delta}
\newcommand{\zg}{\gamma}
\newcommand{\zl}{\lambda}

\newcommand{\deff}{\stackrel{\text{\tiny def}}{=}}

\newtheorem{theorem}{\bf Theorem}
\numberwithin{theorem}{section}

\numberwithin{equation}{section}
\numberwithin{figure}{section}

\author{F. Borghero\footnote{Dip. Matematica e Informatica, Universit\`a di Cagliari,
Via Ospedale 72, 09124 Cagliari, Italy, e-mail: borghero@unica.it}, 
F. Demontis\footnote{Dip. Matematica, Universit\`a di Cagliari,
Viale Merello 92, 09123 Cagliari, Italy, e-mail: fdemontis@unica.it} and 
S. Pennisi \footnote{Dip. Matematica, Universit\`a di Cagliari,
Via Ospedale 72, 09124 Cagliari, Italy, e-mail: spennisi@unica.it}}
\title{Wave Speed in the Macroscopic Extended Model for Ultrarelativistic Gases}

\begin{document}
\date{}
\maketitle
\thispagestyle{plain}

\begin{abstract}
An exact macroscopic extended model for ultrarelativistic gases, with an arbitrary
number of moments, is present in the literature. Here we exploit equations determining wave
speeds for that model. We find interesting results; for example, the whole system for 
their determination can be divided into independent subsystems and some, but not all, wave 
speeds are expressed by rational numbers. Moreover, the extraordinary
property that these wave speeds for the macroscopic model are the same of those
in the kinetic model, is proved.
\end{abstract}

\section{Introduction}\label{sec:1}
The macroscopic extended model, with an arbitrary number of moments, for ultrarelativistic gases
has been introduced in \cite{1} which is the generalization of \cite{2}. It proposes the field equations
\begin{equation}\label{1}
\partial_{\za}A^{\za\za_1\cdots\za_n}=I^{\za_1\cdots\za_n},\quad \text{for $n=0,1,\ldots,N$},
\end{equation}
where $A^{\za\za_1\cdots\za_n}$ and $I^{\za_1\cdots\za_n}$ are symmetric and 
trace-less tensors.
In \cite{1} it is proved that the entropy principle for this system amounts to assuming 
the existence of the symmetric and trace-less Lagrange multipliers 
$\zl_{\za_1\za_2\cdots\za_n}$ and of
an arbitrary function $F(x_0,x_1,\cdots,x_\mu)$, such that
\begin{equation}\label{2}
A^{\za\za_1\cdots\za_m}=\int F_m(\zl,\zl_\mu p^\mu,\cdots,\zl_{\mu_1\cdots\mu_{N}}p^{\mu_1}
\cdots p^{\mu_{N}}) p^\za p^{\za_1}\cdots p^{\za_m}dP,
\end{equation}
where $F_m=\dfrac{\partial F}{\partial X_m},\quad dP=\dfrac{dp^1dp^2dp^3}{p^0}$ and $p^{\mu}$
is the four momentum satisfying the relation $p^{\mu}p_{\mu}=0$ (because the particle mass is 
zero in an ultrarelativistic gas). Moreover, $\zl_{\za_1\cdots \za_n}$ can be taken as independent 
variables. A further condition has been studied  in \cite{3}, while a reader not acquainted with
the general context of Extended Thermodynamics might profit by reading \cite{4, 5, 6}.

In this article we study the wave speeds for the system \eqref{1}. To this end we observe that this 
system, taking into account \eqref{2}, can be written as 
\begin{equation}\label{3}
\sum_{n=0}^N A^{\za\za_1\cdots\za_m\zb_1\cdots\zb_n}\partial_{\za}
\zl_{\zb_1 \cdots \zb_n}=I^{\za_1 \cdots \za_n},
\end{equation}
with
\begin{align}\label{4}
\nonumber &A^{\za\za_1\cdots\za_m\zb_1\cdots \zb_n}\\&=\hspace{-0.2cm}\int F_{m,n}
(\zl,\zl_\mu p^\mu,\cdots,\zl_{\mu_1\cdots\mu_{N}}p^{\mu_1}
\cdots p^{\mu_{N}}) p^\za p^{\za_1}\cdots p^{\za_m}p^{\zb_1}\cdots p^{\zb_n}dP,
\end{align}
where $F_{m,n}=\dfrac{\partial^2 F}{\partial X_m \partial X_n}$. Now recall that 
hyperbolicity for 
the system \eqref{3} in the time-like direction $\xi_{\za}$
(with $\xi_{\za}\xi^{\za}=-1$), in the sense of Friedrichs \cite{7, 8, 9}, 
means that
\begin{enumerate}
\item The system $\xi_{\za}A^{\za\za_1\cdots\za_m\zb_1\cdots \zb_n}\zd\zl_{\zb_1\cdots\zb_n}=0,$ 
in the unknown $d\zl_{\zb_1\cdots\zb_n}$, has only the solution $\zd\zl_{\zb_1\cdots\zb_n}=0$;
\item The system 
\begin{equation}\label{2.5}
\varphi_{\za}A^{\za\za_1\cdots\za_m\zb_1\cdots \zb_n}\zd\zl_{\zb_1\cdots\zb_n}=0,
\end{equation}
where $\varphi_\za=\eta_\za-\zl\xi_\za$, has real eigenvalues $\zl$ and a basis of 
eigenvectors $\zd\zl_{\zb_1\cdots\zb_n}$, for every unit vector $\eta_\za$ such that
$\eta_\za\eta^\za=1,\,\eta_\za\xi^\za=0$.
\end{enumerate}
The eigenvalues $\zl$ are the wave speeds to obtain.
A sufficient condition for hyperbolicity in every time-like direction can be found in \cite{10}.
The hyperbolicity for our system \eqref{3} has already been proved in \cite{1} as a consequence of the so-called 
``convexity requirement''; so it remains to find the values of the eigenvalues. We first
introduce the time-like direction $u_\za=\zg^{-1}\zl_\za$ with $\zg=\sqrt{-\zl_\mu\zl^\mu}$ and the 
projector 
\begin{equation}\label{5}
h_{\za\zb}=g_{\za\zb}+u_\za u_\zb
\end{equation}
into the $3$-dimensional subspace orthogonal to $u_\za$. After that we note that $h^{\za\zb}\varphi_\zb\neq 0$,
otherwise $\varphi_\za=-\phi_\mu u^\mu u_\za$ and the condition $2.$ becomes $1.$ with $\xi_\za=u_\za$,
from which $d\zl_{\zb_1\cdots\zb_n}=0$ because of the fact that $d\zl_{\zb_1\cdots\zb_n}$ is an eigenvector. 
Because $h^{\za\zb}\varphi_\zb\neq 0$, we can now define 
$\varphi=\sqrt{h^{\mu\eta}\varphi_\mu\varphi_\eta}$ and $v_\za=\varphi^{-1}h_\za^\zb\varphi_{\zb}$, so that 
$v_\za$ is a unit vector orthogonal to $u_\za$. Finally, we introduce $K_{\za\zb}=h_{\za\zb}-v_\za v_\zb$
which is the projector onto the $2$-dimensional subspace orthogonal to $u_\za$ and $v_\za$.

In \cite{BDP} we have found some properties satisfied by these projectors and the definition of 
$K_{\zb_1}^{<_2\,\,\zg_1}\cdots K_{\zb_p}^{\zg_p>}$, a tensor such that for every tensor $T^{\zb_1\cdots\zb_p}$
the tensor $T^{<_2\,\, \zg_1\cdots \zg_p>}=K_{\zb_1}^{<_2\,\, \zg_1}\cdots K_{\zb_p}^{\zg_p>} T^{\zb_1\cdots\zb_p}$ 
is equal to the sum of the expression $K_{\zb_1}^{\zg_1}\cdots K_{\zb_p}^{\zg_p}T^{\zb_1\cdots \zb_p}$ 
and a linear combination of its 
$2$-dimensional traces (i.e., $K_{\zb_1\zb_2}\cdots K_{\zb_{2s-1}\zb_{2s}}K_{\zb_{2s+1}}^{\zg_{2s+1}}\cdots 
K_{\zb_p}^{\zg_p}T^{\zb_1\cdots\zb_p}$) and, moreover, the $2$-dimensional trace of $T^{<_2\,\,\zg_1\cdots \zg_p>}$
is equal to zero. In other words, $T^{<_2\,\,\zg_1\cdots \zg_p>}$ is the $2$-dimensional trace-less part of 
$K_{\zb_1}^{\zg_1}\cdots K_{\zb_p}^{\zg_p}T^{\zb_1\cdots \zb_p}$. In Section \ref{sec:2} we will report
some of these results. We recall that, in \cite{BDP}, only the case $N=3$ has been treated, 
while in the present paper we study the general case. However, some details of \cite{BDP} are also general, so 
we report these results here, for the sake of completeness. 
Obviously, the objective of all these considerations is 
to rewrite the system \eqref{2.5} in terms of tensors belonging to our $2$-dimensional subspace.
This result will be 
given in Section \ref{sec:2}. 

The system \eqref{2.5} will not assume a very elegant form, a consequence of its generality.
For this reason in section \ref{sec:3} we will consider the cases $p=N,\, p=N-1$ and $p=N-2$. 
In this way the case $N=2$
will be exhausted. The resulting wave speed, in the reference frame moving along 
with the fluid, are $\zl=0,\,\zl=\frac{1}{5},
\,\zl=\pm 1,\,\zl=\pm\frac{1}{\sqrt{3}}.$ These are expressed by rational numbers, 
except for the last one.

Finally, in Section \ref{sec:4} we will prove the extraordinary property that the 
wave speeds do not depend on the function
$F$, so that we can take $F=e^{-\sum_{i=0}^{N}X_i/K}$ where $K$ is the 
Boltzmann constant; in other words,
the wave speeds for the macroscopic model are the same as those in 
the kinetic model \cite{11}.

\section{On the $2$-dimensional trace-less part of a tensor and other properties of the projectors} \label{sec:2}
Let us begin by exploiting some technical tensorial properties which will be useful to write
the system \eqref{2.5} in terms of $2$-dimensional traceless tensors.

First, let us define the tensor
\begin{align}\label{7.1}
K_{\zb_1}^{<_2\,\,\zg_1}\cdots K_{\zb_p}^{\zg_p>}=\sum_{s=0}^{\left[\frac{p}{2}\right]}
a_s \,K_{(\zb_1\zb_2}\cdots K_{\zb_{2s-1}\zb_{2s}}K_{\zb_{2s+1}}^{(\zg_{2s+1}}\cdots K_{\zb_{p})}^{\zg_{p}}
K^{\zg_1\zg_2}\cdots K^{\zg_{2s-1}\zg_{2s})}\,,
\end{align}
where $a_s=\left(-\frac{1}{4}\right)^s\frac{p}{(p-2s)!}\frac{(p-s-1)!}{s!}$ and from now on the square brackets 
denote the integer parts of the number. 

What does this tensor mean? When we contract it with a generic symmetric tensor $T^{\zb_1\cdots \zb_p}$ we 
obtain that $T^{<_2\,\, \zg_1\cdots \zg_p>}\deff K_{\zb_1}^{<_2\,\,\zg_1}\cdots K_{\zb_p}^{\zg_p>}T^{\zb_1\cdots \zb_p}$
equals the sum of $T^{\zb_1\cdots \zb_p}K_{\zb_1}^{\zg_1}\cdots K_{\zb_p}^{\zg_p}$ (because $a_0=1$) and of a linear
combination of the $2$-dimensional traces of $T^{\zg_1\cdots \zg_p}$. Moreover, the following theorem holds:
\begin{theorem}\label{th:1}
The $2$-dimensional trace of $T^{<_2\,\, \zg_1\cdots \zg_p}$ is zero, or, equivalently 
\begin{equation}\label{7.2}
K_{\zb_1}^{<_2\,\,\zg_1}\cdots K_{\zb_p}^{\zg_p>}K_{\zg_1\zg_2}=0\,\,.
\end{equation}
\end{theorem}
This theorem has been proved in \cite{BDP} as are the other theorems of this section. By means of
Theorem \ref{th:1} it is now natural to call $T^{<_2\,\, \zg_1\cdots\zg_p>}$ the $2$-dimensional trace-less 
part of $T^{\zg_1\cdots\zg_p}$. Then \eqref{7.1} gives the $2$-dimensional trace-less part of $K_{\zb_1}^{(\zg_1}\cdots 
K_{\zb_p}^{\zg_p)}$; it will be useful for the sequel to have a sort of inversion of this relation. It is provided by
\begin{theorem}\label{th:2}
The following identity holds:
\begin{align}\label{8.1}
K_{\za_1}^{(\zb_1}\cdots K_{\za_r}^{\zb_r)}=\sum_{s=0}^{\left[\frac{r}{2}\right]}b_{r,s}\,\,K_{(\za_1\za_2}\cdots
K_{(\za_{2s-1}\za_{2s}}K^{(\zb_{2s+1}}_{<_2\,\,\za_{2s+1}}K^{\zb_r}_{\za_r>}K^{\zb_1\zb_2}\cdots K^{\zb_{2s-1}\zb_{2s})}\,,
\end{align}
with 
\begin{align}\label{8.1_2}
b_{r,s}=\frac{r!}{(r-s)!}\frac{1}{(2s)!!}\frac{(2r-4s)!!}{(2r-2s)!!}.
\end{align}
\end{theorem}
Also the proof of this theorem is reported in \cite{BDP}. We conclude this list of theorems with 
\begin{theorem}\label{th:3}
The following identity holds:
\begin{align}\label{8.2}
\nonumber & K_{\za_1}^{<_2\,\,\zg_1}\cdots K_{\za_p}^{\zg_p>}K^{(\za_1\za_2}\cdots K^{\za_{2s-1}\za_{2s}}
V^{\za_{2s+1}}\cdots V^{\za_{2s+c}} U^{\za_{2s+c+1}}\cdots U^{\za_{2s+c+d})}
\\ \nonumber=& \dfrac{(2s)!!}{(2s-2p)!!}\dfrac{(2s+c+d-p)!}{(2s+c+d)!}K_{\za_1}^{<_2\,\,\zg_1}\cdots K_{\za_p}^{\zg_p>}
K^{\za_1\underline{\za_{p+1}}}K^{\za_2\underline{\za_{p+2}}}\cdots K^{\za_p\underline{\za_{2p}}}\\ \cdot& 
K^{\underline{\za_{2p+1}}\underline{\za_{2p+2}}}\cdots K^{\underline{\za_{2s-1}}\underline{\za_{2s}}}
V^{\underline{\za_{2s+1}}}\cdots V^{\underline{\za_{2s+c}}}U^{\underline{\za_{2s+c+1}}}\cdots U^{\underline{\za_{2s+c+d}}}\,,
\end{align}
\end{theorem}
where the indices underlined denote a symmetrization over those indices. See \cite{BDP} for the proof of this
theorem.

In \cite{BDP} we have also proved that the equation
\begin{align}\label{I.7}
X^{\zg_1\cdots \zg_p}_{(h, k)}=K_{\zb_1}^{<_2\,\,\zg_1}\cdots K_{\zb_p}^{\zg_p>}U_{\zb_{p+1}}
\cdots U_{\zb_{p+h}}V_{\zb_{p+h+1}}\cdots V_{\zb_{p+h+k}}d\zl^{\zb_1\cdots \zb_n}\,\, ,
\end{align}
represents an invertible transformation from the variables
$d\zl_{\zb_1\cdots \zb_n}$ (with zero $4$-dimensional trace) to $X_{(h,k)}^{\zg_1\cdots\zg_p}$
(belonging to the subspace orthogonal to $U_{\zg_1}$ and $V_{\zg_1}$ and with zero $2$-dimensional trace) 
and the condition that $d\zl_{\zb_1\cdots\zb_n}$ has zero $4$-dimensional trace
is ``translated'' into the fact that $X^{\zg_1\cdots \zg_p}_{(h, k)}$ has $2$-dimensional zero trace.
We note that the index $h$ denotes how many contractions with $U_{\cdots}$ are involved, while the index 
$k$ denotes how many contractions with $V_{\cdots}$ are present.

Therefore, we can now consider $X_{(h,\,k)}^{\zg_1\cdots \zg_p}$ as unknowns, instead of $d\zl_{\zb_1\cdots \zb_n}$.
Obviously, the invertibility above mentioned holds for any symmetric tensor $d\zl_{\za_1\cdots \za_n}$
with zero $4$-dimensional trace; consequently, it holds also with $\varphi_\za A^{\za\zb_1\cdots \zb_m\za_1\cdots\za_n}
d\zl_{\za_1\cdots \za_n}$ instead of $d\zl_{\za_1\cdots \za_n}$. In other words, we have that if 
\begin{align}\label{C.3} 
U_{\zb_1}\cdots U_{\zb_a}V_{\zb_{a+1}}\cdots V_{\zb_{a+b}}K^{<_2\,\,\zg_{a+b+1}}_{\zb_{a+b+1}}
\cdots K^{\zg_m>}_{\zb_m}\sum_{n=0}^N \varphi_\za A^{\za\zb_1\cdots \zb_m\za_1\cdots\za_n}
d\zl_{\za_1\cdots \za_n}=0
\end{align}
for all $a,\,b$ such that $a+b\leq m$, then also \eqref{2.5} will be satisfied. The converse is 
trivial. Consequently, \eqref{C.3} is the new system to be used 
to obtain the unknowns $X_{(h,\,k)}^{\zg_1\cdots\zg_p.}$
But it is better to rewrite \eqref{C.3} as 
\begin{align}\label{E.1}
\nonumber & \sum_{n=0}^{N}\left(-\varphi_\mu U^\mu U_\za+\varphi V_\za\right)
K^{<_2\,\,\zg_1}_{\zb_1}\cdots K^{\zg_p\,\,>}_{\zb_p}U_{\zb_{p+1}}\cdots U_{\zb_{p+a}}
V_{\zb_{p+a+1}}\cdots V_{\zb_{p+a+b}}\\& \cdot A^{\za\zb_1\cdots \zb_m\za_1\cdots \za_n}d\zl_{\za_1\cdots \za_n}=0\,\,,
\end{align}
with $p=0,\ldots,m$; $a+b=m-p$.

It is interesting to see that, from \cite{BDP}, it is not necessary to substitute the relation linking
$d\zl_{\za_1\cdots \za_n}$ to $X_{(h,\,k)}^{\zg_1\cdots \zg_p}$ in this equation, since
 it will be a natural consequence of the 
kinetic expression \eqref{4}. The effective steps can be found in \cite{BDP}, so we limit ourselves
to report the result here, i.e.,
\begin{align}\label{7}
\nonumber & -\varphi_\mu u^{\mu} \sum_{n=p+b-2\left[\frac{b}{2}\right]}^{N}\sum_{r=\left[\frac{2p+b+1}{2}\right]}
^{\left[\frac{m+n-a}{2}\right]}\sum_{s=p}^{\left[\frac{2r-b}{2}\right]}\sum_{T=0}^{s-p} \frac{4\pi}{2r+1}
\begin{pmatrix}m+n-a\\2r\end{pmatrix}\cdot (-1)^{m+n-a}\begin{pmatrix}r\\s\end{pmatrix}\\\nonumber & 
\cdot\int_0^{+\infty}F_{m,\,n}(\zl,\rho\zg)\rho^{m+n+2}d\rho \,\frac{(2r-2s)!}{(2r-2s-b)!}\,\frac{(m+n-a-b-p)!}{(m+n-a)!}
\,\frac{(2s)!!}{(2s-2p)!!}\\\nonumber &\cdot \begin{pmatrix}s-p\\T\end{pmatrix}(-1)^T\,\, 
X^{\zg_1\cdots \zg_p}_{(n-p-2r+b+2s-2T, 2r-b-2s+2T)}+\\\nonumber & 
+\varphi \sum_{n=p+b+1-2\left[\frac{b+1}{2}\right]}^{N}\sum_{r=\left[\frac{2p+b+2}{2}\right]}
^{\left[\frac{m+n-a+1}{2}\right]}\sum_{s=p}^{\left[\frac{2r-b-1}{2}\right]}\sum_{T=0}^{s-p} \frac{4\pi}{2r+1}
\begin{pmatrix}m+n+1-a\\2r\end{pmatrix}\cdot (-1)^{m+n+1-a}\begin{pmatrix}r\\s\end{pmatrix}\\\nonumber & 
\cdot\int_0^{+\infty}F_{m,\,n}(\zl,\rho\zg)\rho^{m+n+2}d\rho \,\frac{(2r-2s)!}{(2r-2s-b-1)!}\,\frac{(m+n-a-b-p)!}{(m+n+1-a)!}
\,\frac{(2s)!!}{(2s-2p)!!}\\ &\cdot \begin{pmatrix}s-p\\T\end{pmatrix}(-1)^T\,\, 
X^{\zg_1\cdots \zg_p}_{(n-p-2r+b+2s-2T+1, 2r-b-2s+2T-1)}=0\, \, .
\end{align}

To state the system \eqref{7} we have to choose
a value of $p$ such that $0\leq p\leq N$; after that we have to write \eqref{7} for every $m$ such that 
\begin{equation}\label{4.2}
p\leq m\leq N
\end{equation}
and for every 
couple $(a,b)$ such that $a+b=m-p$. In this way a set of equations is obtained which are linear combinations of tensors
$X^{\zg_1\cdots \zg_p}_{(h, k)}$, all of the same order $p$; moreover, the coefficients of these linear combinations are
scalar functions. In other words, the wave speeds are obtained by simply imposing that the matrix of these scalar coefficients
is singular. Doing this for every value of $p$, we obtain all the wave speeds. As a result we have proved that the whole
system, for the determination of the wave speeds, can be divided into some independent subsystems for each given value of $p$.
We note also that for every fixed value of $p$, from \eqref{2.5} only those with $m\geq p$ have to be considered in the system
\eqref{7}; moreover, in this system we have $n\geq p$ so that, from the initial unknowns $\zd\zl_{\zb_1\cdots\zb_n}$ only those with
$n\geq p$ are present in the system \eqref{7}.

\section{The wave speeds for some values of $p$}\label{sec:3}

\subsection{The wave speeds for $p=N$}\label{subsec:4.1}
In this case, from \eqref{4.2}, we have $m=N.$ In other words, only the 
last equation of system \eqref{2.5} contributes to the system \eqref{7}.
Moreover, $a+b=m-p$ implies $a=b=0$. As a consequence, we see that \eqref{7}
becomes
$$-\varphi_\mu U^\mu \frac{4\pi}{2N+1}G_{N,\,N}\frac{N!}{(2N)!}(2N)!!\,
X_{(0,0)}^{\zg_1\ldots \zg_N}=0\,,$$
because, in the coefficient of $\varphi$ the summation $\displaystyle\sum_{n=N+1}^N$
appears, so that this coefficient is zero. Moreover, we have defined 
$G_{m,\,n}=\int_0^\infty F_{m,\,n}(\zl,\,\rho\zg)\rho^{m+n+2}d\rho.$ 
Consequently,
we have $\varphi_\mu U^\mu=0$ as unique eigenvalue. It corresponds 
to the waves moving along with the fluid, because in the case $\xi_\mu=U_\mu$
we have $0=\varphi_\mu U^\mu=(-\zl U_\mu+\eta_\mu)U^\mu=\zl.$ In other words, 
the wave speed is zero. 

We also note that, $\varphi_\mu U^\mu=0$ does not depend on $F_{m,\,n}$,
so it is the same as for the kinetic model \cite{11}. 
 
\subsection{The wave speeds for $p=N-1$}\label{subsec:4.2}

From \eqref{4.2} we obtain $m=N-1$ or $m=N.$ In other words,
only the last two equations of the system \eqref{2.5} contribute 
to the system \eqref{7}. 

If $m=N-1$, from $a+b=m-p$ it follows $a=0,\,b=0$.
If $m=N$, from the same equation we obtain $a+b=1,$ i.e., $a=1,\,b=0$
or $a=0,\,b=1$. In this way we obtain three equations,
\begin{align}\label{R.2_1}
&-\varphi_\mu U^\mu \frac{4\pi}{(2N-1)!}(N-1)!(2N-2)!!\,G_{N-1,\,N-1}\,
X_{(0,0)}^{\zg_1\ldots \zg_{N-1}}+G_{N-1,\,N}\\&\nonumber \cdot 
\big[\varphi_\mu U^\mu \frac{4\pi}{(2N-1)!}N!(2N-2)!!
X_{(1,0)}^{\zg_1\ldots \zg_{N-1}}+\varphi\frac{8\pi}{(2N+1)!}NN!(2N-2)!!
X_{(0,1)}^{\zg_1\ldots \zg_{N-1}} \big]=0\,,
\end{align}
\begin{align}\label{R.2_2}
&-\varphi_\mu U^\mu \frac{4\pi}{(2N-1)!}(N-1)!(2N-2)!!\,G_{N,\,N-1}\,
X_{(0,0)}^{\zg_1\ldots \zg_{N-1}}+G_{N,\,N}\\&\nonumber \cdot
\big[\varphi_\mu U^\mu \frac{4\pi}{(2N-1)!}N!(2N-2)!!
X_{(1,0)}^{\zg_1\ldots \zg_{N-1}}+\varphi\frac{8\pi}{(2N+1)!}NN!(2N-2)!!
X_{(0,1)}^{\zg_1\ldots \zg_{N-1}}\big]=0\,,
\end{align}
\begin{align}\label{R.2_3}
&\varphi\,\frac{8\pi(N-1)!N}{(2N+1)!}(2N-2)!!\,G_{N,\,N-1}\,
X_{(0,0)}^{\zg_1\ldots \zg_{N-1}}+G_{N,\,N}\\&\nonumber \cdot
\big[-\varphi\frac{8\pi N(N)!}{(2N+1)!}(2N-2)!!
X_{(1,0)}^{\zg_1\ldots \zg_{N-1}}-\varphi_\mu U^\mu\frac{8\pi}{(2N+1)!}
NN!(2N-2)!!X_{(0,1)}^{\zg_1\ldots \zg_{N-1}}\big]=0\,,
\end{align}
respectively. Now we note that \eqref{R.2_1} and \eqref{R.2_2} can be considered as two 
equations in the two unknowns
\begin{align*}
& -\varphi_\mu U^\mu \frac{4\pi}{(2N-1)!}(N-1)!(2N-2)!!\,
X_{(0,0)}^{\zg_1\ldots \zg_{N-1}},\quad\text{and}\\
&\varphi_\mu U^\mu \frac{4\pi}{(2N-1)!}N!(2N-2)!!\,X_{(1,0)}^{\zg_1
\ldots \zg_{N-1}}+
\varphi\frac{8\pi}{(2N+1)!}NN!(2N-2)!!\,X_{(0,1)}^{\zg_1
\ldots \zg_{N-1}}\,,
\end{align*}
where
\begin{align}\label{31.1}
\begin{pmatrix}G_{N-1,\,N-1} & G_{N-1,\,N}\\G_{N,\,N-1} & G_{N,\,N}\end{pmatrix}
\end{align}
is the coefficient matrix. But, in the particular case $\varphi=0,$ we would obtain 
only the zero solution for the equations \eqref{R.2_1}-\eqref{R.2_2} and this for
condition $1.$ of hyperbolicity; therefore the matrix \eqref{31.1}
is nonsingular and $G_{N,\,N}\neq 0,$ so that the two unknowns are zero, that is 
\begin{subequations}\label{5.1}
\begin{align}
&\varphi_\mu U^\mu\,X_{(0,\,0)}^{\zg_1,\ldots,\zg_{N-1}}=0\,,\\
&\varphi_\mu U^\mu\,X_{(1,\,0)}^{\zg_1,\ldots,\zg_{N-1}}+2\varphi\,\frac{N}{2N(2N+1)}
X_{(0,\,1)}^{\zg_1,\ldots,\zg_{N-1}}=0\,.
\end{align}
\end{subequations}
Also \eqref{R.2_3} has to be imposed. Now, if $\varphi_\mu U^\mu=0$,
these three equations
become 
\begin{subequations}\label{31.3}
\begin{align}
& X_{(0,\,1)}^{\zg_1,\ldots,\zg_{N-1}}=0\,,\\
& G_{N,\,N-1}X_{0,\,0}^{\zg_1,\ldots,\zg_{N-1}}-G_{N,\,N}
X_{(1,\,0)}^{\zg_1,\ldots,\zg_{N-1}}=0\,,
\end{align}
\end{subequations}
so that there is a free unknown. Consequently, $\varphi_\mu U^\mu=0$ is an eigenvalue.
If $\varphi_\mu U^\mu\neq 0$, \eqref{5.1} and \eqref{R.2_3} can be written as
\begin{align*}
& X_{0,\,0}^{\zg_1,\ldots,\zg_{N-1}}=0\,,\\
& \begin{pmatrix} \varphi_\mu U^\mu & \frac{\varphi}{2N+1}\\ 
\frac{\varphi}{2N+1} & \varphi_\mu U^\mu \end{pmatrix}
\begin{pmatrix} X_{(1,\,0)}^{\zg_1,\ldots,\zg_{N-1}}\\X_{(0,\,1)}^{\zg_1,
\ldots,\zg_{N-1}} \end{pmatrix}=
\begin{pmatrix}0\\0\end{pmatrix}\,,
\end{align*}
so that the eigenvalues have to be determined from the equation 
\begin{equation}\label{5.2}
(\varphi_\mu U^\mu)^2-\frac{\varphi^2}{(2N+1)^2}=0\,.
\end{equation}
Thus \eqref{5.2} does not depend on $G_{M,\,N}$ and hence it is 
the same as in the kinetic case.
Instead, the eigenvectors defined by \eqref{31.3} depend on 
$G_{M,\,N}$, so that they are not the same as in the kinetic case. In particular,
if $\xi_\mu=U_\mu$, from which $\varphi_\mu U^\mu=\zl$ and $\varphi=1$,
the wave speeds are $\zl=\pm\frac{1}{2N+1}$ which are rational numbers satisfying
 $|\zl|<1,$ which means that the wave
velocity does not exceed the speed of light, as expected.
Moreover, our results is consistent
with \cite{12}.

\subsection{The wave speeds for $p=N-2$}\label{subsec:4.3}

This part is important because, if $N=2$, it completes the set of wave speeds
for the $14$ moments model. From \eqref{4.2} we now have $m=N-2$ or $m=N-1$
or $m=N;$ more precisely, we have to write the \eqref{7} with
\begin{itemize}
\item $m=N-2,\quad a=0, \quad b=0,$
\item $m=N-1,\quad a=1, \quad b=0,$
\item $m=N,\quad\quad \,\,\,\,\,a=2, \quad b=0,$
\item $m=N-1,\quad a=0, \quad b=1,$
\item $m=N,\quad\quad \,\,\,\,\, a=1, \quad b=1,$
\item $m=N,\quad\quad \,\,\,\,\, a=0, \quad b=2.$
\end{itemize}
In this way we obtain the system
\begin{subequations}\label{33.1}
\begin{align}
 & G_{N-2,\,N-2}\,X_1^{\zg_1\ldots\zg_{N-2}}+G_{N-2,\,N-1}\,X_2^{\zg_1\ldots\zg_{N-2}}+
G_{N-2,\,N}\,X_3^{\zg_1\ldots\zg_{N-2}}=0\,, \label{33.1a}\\
 & G_{N-1,\,N-2}\,X_1^{\zg_1\ldots\zg_{N-2}}+G_{N-1,\,N-1}\,X_2^{\zg_1\ldots\zg_{N-2}}+
G_{N-1,\,N}\,X_3^{\zg_1\ldots\zg_{N-2}}=0\,,\label{33.1b}\\
 & G_{N,\,N-2}\,X_1^{\zg_1\ldots\zg_{N-2}}+G_{N,\,N-1}\,X_2^{\zg_1\ldots\zg_{N-2}}+
G_{N,\,N}\,X_3^{\zg_1\ldots\zg_{N-2}}=0\,,\label{33.1c}\\
 & G_{N-1,\,N-2}\,Y_1^{\zg_1\ldots\zg_{N-2}}+G_{N-1,\,N-1}\,Y_2^{\zg_1\ldots\zg_{N-2}}+
G_{N-1,\,N}\,Y_3^{\zg_1\ldots\zg_{N-2}}=0\,,\label{33.1d}\\
 & G_{N,\,N-2}\,Y_1^{\zg_1\ldots\zg_{N-2}}+G_{N,\,N-1}\,Y_2^{\zg_1\ldots\zg_{N-2}}+
G_{N,\,N}\,Y_3^{\zg_1\ldots\zg_{N-2}}=0\,,\label{33.1e}\\
& G_{N,\,N-2}\,Z_1^{\zg_1\ldots\zg_{N-2}}+G_{N,\,N-1}\,Z_2^{\zg_1\ldots\zg_{N-2}}+
G_{N,\,N}\,Z_3^{\zg_1\ldots\zg_{N-2}}=0\,,\label{33.1f}
\end{align}
\end{subequations}
where
\begin{subequations}\label{34.1}
\begin{align}
& X_1^{\zg_1\ldots\zg_{N-2}}=-\varphi_\mu U^\mu\frac{4\pi}{(2N-3)!}(N-2)!(2N-4)!!
\,X_{(0,\,0)}^{\zg_1\ldots\zg_{N-2}}\,,\label{34.1a}\\
\nonumber & X_2^{\zg_1\ldots\zg_{N-2}}=\varphi_\mu U^\mu\frac{4\pi}{(2N-3)!}(N-1)!(2N-4)!!
\,X_{(1,\,0)}^{\zg_1\ldots\zg_{N-2}}\\  &+\varphi\, \frac{4\pi}{(2N-1)!}2(N-1)(N-1)!(2N-4)!!
\,X_{(0,\,1)}^{\zg_1\ldots\zg_{N-2}}\,,\label{34.1b}\\
\nonumber& X_3^{\zg_1\ldots\zg_{N-2}}=-\varphi_\mu U^\mu\frac{4\pi}{(2N-3)!}\frac{1}{2}N!(2N-4)!!
\,X_{(2,\,0)}^{\zg_1\ldots\zg_{N-2}}\\\nonumber&\cancel{-\varphi_\mu U^\mu \frac{4\pi}{(2N-1)!}(N-1)N!(2N-4)!!
\,X_{(0,\,2)}^{\zg_1\ldots\zg_{N-2}}}\\\nonumber&-\varphi_\mu U^\mu \frac{4\pi}{(2N-1)!}\frac{1}{2}N!(2N-2)!!
\,X_{(2,\,0)}^{\zg_1\ldots\zg_{N-2}}\\\nonumber&\cancel{+\varphi_\mu U^\mu \frac{4\pi}{(2N-1)!}\frac{1}{2}N!(2N-2)!!
\,X_{(0,\,2)}^{\zg_1\ldots\zg_{N-2}}}\\&-\varphi\,\frac{4\pi}{(2N+1)!}(2N-1)(N-1)2\frac{N!}{(2N-1)!}(2N-4)!!
\,X_{(1,\,1)}^{\zg_1\ldots\zg_{N-2}}\,,\label{34.1c}\\
& Y_1^{\zg_1\ldots\zg_{N-2}}=\varphi\,\frac{4\pi}{(2N-1)!}2N(N-1)!(2N-4)!!
\,X_{(0,\,0)}^{\zg_1\ldots\zg_{N-2}}\,,\label{34.1d}\\
\nonumber & Y_2^{\zg_1\ldots\zg_{N-2}}=-\varphi_\mu U^\mu\frac{4\pi}{(2N-1)!}(N-1)2(N-1)!(2N-4)!!
\,X_{(0,\,1)}^{\zg_1\ldots\zg_{N-2}}\\ &-\varphi\,\frac{4\pi}{(2N-1)!}2(N-1)(N-1)!(2N-4)!!
\,X_{(1,\,0)}^{\zg_1\ldots\zg_{N-2}}\,,\label{34.1e}
\end{align}
\begin{align}
\nonumber & Y_3^{\zg_1\ldots\zg_{N-2}}=\varphi_\mu U^\mu\frac{4\pi}{2N+1}\frac{N!}{(2N-1)!}(2N-1)(N-1)2(2N-4)!!
\,X_{(1,\,1)}^{\zg_1\ldots\zg_{N-2}}\\\nonumber&+\varphi \frac{4\pi}{(2N-1)!}(N-1)N!(2N-4)!!
\,X_{(2,\,0)}^{\zg_1\ldots\zg_{N-2}}\\\nonumber&+6\varphi\, \frac{4\pi}{(2N+1)!}N(N-1)N!(2N-4)!!
\,X_{(0,\,2)}^{\zg_1\ldots\zg_{N-2}}\\&+\varphi\,\frac{4\pi}{(2N+1)!}NN!(2N-2)!!
\,X_{(2,\,0)}^{\zg_1\ldots\zg_{N-2}}-\varphi\,\frac{4\pi}{(2N+1)!}NN!(2N-2)!!
\,X_{(0,\,2)}^{\zg_1\ldots\zg_{N-2}}\,,\label{34.1f}\\
 & Z_1^{\zg_1\ldots\zg_{N-2}}=-\varphi_\mu U^\mu\,\frac{4\pi}{(2N-1)!}2(N-1)!(2N-4)!!
\,X_{(0,\,0)}^{\zg_1\ldots\zg_{N-2}}\,,\label{34.1g}\\
\nonumber & Z_2^{\zg_1\ldots\zg_{N-2}}=-\varphi_\mu U^\mu\frac{4\pi}{(2N-1)!}2(N-1)(N-1)!(2N-4)!!
\,X_{(1,\,0)}^{\zg_1\ldots\zg_{N-2}}\\&+\varphi\,\frac{4\pi}{(2N+1)!}12(N-1)N!(2N-4)!!
\,X_{(0,\,1)}^{\zg_1\ldots\zg_{N-2}}\,,\label{34.1h}\\
\nonumber & Z_3^{\zg_1\ldots\zg_{N-2}}=-\varphi_\mu U^\mu\frac{4\pi}{(2N-1)!}(N-1)N!(2N-4)!!
\,X_{(2,\,0)}^{\zg_1\ldots\zg_{N-2}}\\\nonumber&-\varphi_\mu U^\mu \frac{4\pi}{(2N+1)!}6N!N(N-1)(2N-4)!!
\,X_{(0,\,2)}^{\zg_1\ldots\zg_{N-2}}\\\nonumber&-\varphi_\mu U^\mu\, \frac{4\pi}{(2N+1)!}NN!(2N-2)!!
\,X_{(2,\,0)}^{\zg_1\ldots\zg_{N-2}}\\\nonumber&+\varphi_\mu U^\mu\,\frac{4\pi}{(2N+1)!}NN!(2N-2)!!
\,X_{(0,\,2)}^{\zg_1\ldots\zg_{N-2}}\\&-\varphi\,\frac{4\pi}{(2N+1)!}12N(N-1)N!(2N-4)!!
\,X_{(1,\,1)}^{\zg_1\ldots\zg_{N-2}}\,.\label{34.1i}
\end{align}
\end{subequations}
Now, \eqref{33.1a}-\eqref{33.1c} give $X_1^{\zg_1\ldots\zg_{N-2}}=0,\,
X_2^{\zg_1\ldots\zg_{N-2}}=0$ and $X_3^{\zg_1\ldots\zg_{N-2}}=0.$ 

If $\varphi_\mu U^\mu=0,$ 
\eqref{34.1a} becomes an identity so that we have five equations in the six unknowns 
$X_{(0,\,0)}^{\zg_1\ldots\zg_{N-2}},\,X_{(0,\,1)}^{\zg_1\ldots\zg_{N-2}},\,
X_{(1,\,0)}^{\zg_1\ldots\zg_{N-2}}$,
$X_{(2,\,0)}^{\zg_1\ldots\zg_{N-2}},\,X_{(1,\,1)}^{\zg_1\ldots\zg_{N-2}},$ and 
$X_{(0,\,2)}^{\zg_1\ldots\zg_{N-2}};$ consequently $\varphi_\mu U^\mu=0$ gives a
wave speed.

If $\varphi_\mu U^\mu \neq0,$ from \eqref{34.1a} we have 
$X_{(0,\,0)}^{\zg_1\ldots\zg_{N-2}}=0.$ From
\eqref{34.1d}, \eqref{34.1g}, and \eqref{34.1h} 
we obtain $Y_1^{\zg_1\ldots\zg_{N-2}}=0,$ and 
$Z_1^{\zg_1\ldots\zg_{N-2}}=0;$ after that 
\eqref{33.1d} and \eqref{33.1e} give 
$Y_2^{\zg_1\ldots\zg_{N-2}}=0$ and $Y_3^{\zg_1\ldots\zg_{N-2}}=0$.
Consequently, we have the equations
\begin{align*}
 X_2^{\zg_1\ldots\zg_{N-2}}=0\,,\,\,  X_3^{\zg_1\ldots\zg_{N-2}}=0\,,
\,\,Y_2^{\zg_1\ldots\zg_{N-2}}=0\,,\,\,  Y_3^{\zg_1\ldots\zg_{N-2}}=0\,,
\end{align*}
together with \eqref{33.1f}, to determine 
$X_{(0,\,1)}^{\zg_1\ldots\zg_{N-2}}$, $X_{(1,\,0)}^{\zg_1\ldots\zg_{N-2}},\,
X_{(0,\,2)}^{\zg_1\ldots\zg_{N-2}}$,
$X_{(2,\,0)}^{\zg_1\ldots\zg_{N-2}}$, $X_{(1,\,1)}^{\zg_1\ldots\zg_{N-2}}.$
Now $X_2^{\zg_1\ldots\zg_{N-2}}=0$ and $Y_2^{\zg_1\ldots\zg_{N-2}}=0$ is a subsystem
of two equations in the unknowns $X_{(1,\,0)}^{\zg_1\ldots\zg_{N-2}}$, 
$X_{0,\,1)}^{\zg_1\ldots\zg_{N-2}}.$ If $(\varphi_\mu U^\mu)^2-\frac{1}{2N-1}\varphi^2=0,$
one of these equations is a consequence of the other, so that we have four equations
to determine five unknowns. Therefore, we have obtained another wave speed. Also in this case
it is smaller than the speed of light, but it is not a rational number.

If $(\varphi_\mu U^\mu)^2-\frac{1}{2N-1}\varphi^2 \neq 0,$ then the equations 
\begin{align*}
 X_2^{\zg_1\ldots\zg_{N-2}}=0\,,\qquad Y_2^{\zg_1\ldots\zg_{N-2}}=0\,,
\end{align*}
give $X_{(1,\,0)}^{\zg_1\ldots\zg_{N-2}}=0$ and 
$X_{0,\,1)}^{\zg_1\ldots\zg_{N-2}}=0$. From \eqref{34.1i} it
follows that $Z_2^{\zg_1\ldots\zg_{N-2}}=0,$ so that \eqref{33.1f}
becomes $Z_3^{\zg_1\ldots\zg_{N-2}}=0.$ Now we are left with the equations 
\begin{align*}
X_3^{\zg_1\ldots\zg_{N-2}}=0\,,\,Y_3^{\zg_1\ldots\zg_{N-2}}=0\,,\,\,Z_3^{\zg_1\ldots\zg_{N-2}}=0
\end{align*}
to determine the unknowns $X_{(0,\,2)}^{\zg_1\ldots\zg_{N-2}},\,
X_{(2,\,0)}^{\zg_1\ldots\zg_{N-2}},\,X_{(1,\,1)}^{\zg_1\ldots\zg_{N-2}}.$
So the last wave speeds are obtained when the coefficent matrix is singular, i.e., when
\begin{align*}
\begin{vmatrix}
0 & -2N(2N+1)\varphi_\mu U^\mu & -(2N-1)\varphi\\
\varphi & (2N+2)\varphi & (2N-1)\varphi_\mu U^\mu\\
-\varphi_\mu U^\mu & -(2N+2)\varphi_\mu U^\mu & -3\varphi
\end{vmatrix}=0\,,
\end{align*}
where we have dropped some factors. The solutions are $\varphi_\mu U^\mu=0$ and 
$(\varphi_\mu U^\mu)^2-\frac{3}{2N-1}\varphi^2=0.$

To conclude this subsection we report the wave speeds found, i.e.,
$$\varphi_\mu U^\mu=0,\quad (\varphi_\mu U^\mu)^2=\frac{1}{2N-1}\varphi^2,\quad
(\varphi_\mu U^\mu)^2=\frac{3}{2N-1}\varphi^2\,,$$
where $\frac{1}{2N-1}\leq 1,\,\frac{3}{2N-1}\leq 1$ (the last one 
inequality holds only for $N\geq 2.$ On the the other hand, if $N<2,$ the case $p=N-2$ has 
not to be considered).

In the particular case $N=2$, which corresponds to the $14$ momements model, we have found 
all the wave speeds, namely
$$\varphi_\mu U^\mu=0,\quad(\varphi_\mu U^\mu)^2=\frac{1}{25}\varphi^2,\quad
(\varphi_\mu U^\mu)^2=\frac{1}{3}\varphi^2,\quad (\varphi_\mu U^\mu)^2=\varphi^2,$$
where the last equation gives the speed of light.

We note that, in any case, we have found something like
\begin{align}\label{38.1}
(\varphi_\mu U^\mu)^2=k\varphi^2,\,\,\text{with $0\leq k \leq 1$.}
\end{align}
On the other hand, the wave speed $\zl$ is defined by 
\begin{align}\label{38.2}
\big[(-\zl\xi_\mu+\eta_\mu)U^\mu)\big]^2=k h^{\mu v}(-\zl\xi_\mu+\eta_\mu)
(-\zl\xi_v+\eta_v)\,,
\end{align}
as in the definition of hyperbolicity reported above. Obviously,
if $\xi_\mu=U_\mu,$ then \eqref{38.2} becomes $\zl^2=k$; in other
words, $\pm\sqrt{k}$ is the wave speed in the reference frame moving along with the fluid.
So, the following question arises immediately: If $0\leq k \leq 1,$ will $\zl$
satisfy also the condition $|\zl|\leq 1$? The answer is affirmative and this will be proved 
in the next subsection.

\subsection{The wave speed in every time-like direction $\xi_\mu$}

The aim of this subsection is to prove that, for $0\leq k \leq 1$, the solutions
$\zl$ of \eqref{38.2} will be such that $-1\leq \zl\leq 1$ for every $\xi_\mu,\,\eta_\mu$
satisfying the conditions
$$\xi_\mu\xi^\mu=-1,\quad \xi_\mu\eta^\mu=0,\quad \eta_\mu\eta^\mu=1.$$
Let us consider the reference frame where $\xi_\mu=(1,\,0,\,0,\,0)$ and
$\eta_\mu=(0,\,1,\,0,\,0)$. In this frame \eqref{38.2} can be written as
$$(-\zl U^0+U^1)^2=k(1-\zl^2)+k(-\zl U^0+U^1)^2,$$ i.e., the wave speeds $\zl$
are the solutions of $f(\zl)=0,$ with
$$f(\zl)=\zl^2\big[(U^0)^2(1-k)+k\big]-2\zl U^0U^1(1-k)+(U^1)^2(1-k)-k,$$
where we note that the coefficient of $\zl^2$ is positive because $0\leq k \leq 1$.
The first question is: Are the roots of $f(\zl)$ real? The answer is affirmative
because $$\frac{\Delta}{4}=k(1-k)\big[(U^0)^2-(U^1)^2\big]+k^2=k+k(1-k)
\big[(U^2)^2+(U^3)^2\big]\geq 0,$$ where we have used the property 
$-(U^0)^2+\displaystyle\sum_{i=1}^3 (U^i)^2=-1$.
Moreover, we easily see that
\begin{align*} 
& f(\pm 1)=(1-k)(U^0\mp U^1)^2\geq 0\,,\\
& \frac{1}{2}f'(\pm 1)=\pm\frac{1}{2} (U^0)^2(1-k)\pm\frac{1}{2}(U^0)^2(1-k)-U^0U^1(1-k)\pm k\,.
\end{align*}
Subsituting $(U^0)^2=1+\sum_{i=1}^3 (U^i)^2$ in the second term
with $(U^0)^2$, it follows that
$$\frac{1}{2}f'(\pm 1)=\pm\frac{1}{2}(1-k)(U^0\mp U^1)^2\pm\frac{1}{2}
\big[(U^2)^2+(U^3)^2\big](1-k)\pm\frac{1}{2}(1+k),$$ so that
$f'(1)>0$ and $f'(-1)<0.$ From $f(1)\geq 0$ and $f'(1)>0$ we obtain $\zl\leq 1,$ 
while from $f(-1)\geq 0,$ and $f'(-1)<0$ we obtain $\zl\geq -1,$ as expected. 

\section{Independence of the wave speeds on $F_{m,\,n}$}\label{sec:4}

Let us now prove that the wave speeds do not depend on $G_{m,\,n}$ and, hence,
on $F_{m,\,n}$. This fact implies that they are the same as for the kinetic 
model, where
\begin{subequations}
\begin{align}\label{40.1}
F_{m,\,n}=\frac{1}{k^2}e^{-\frac{1}{k}(\zl+\zl_\mu p^\mu)},\quad\text{and}\,\quad
G_{m,\,n}=\int_0^\infty \,\frac{1}{k^2}e^{-\frac{1}{k}(\zl+\rho\zg)}\rho^{m+n+2}d\rho.
\end{align}
\end{subequations}
To obtain this result, let us rewrite \eqref{7} with $a=m-p-b.$
We note that now the index $m$ is present only in $G_{m,\,n}.$ Therefore,
\eqref{7} can be written as 
\begin{align}\label{41.1}
\sum_{n=p}^N G_{m,\,n}Y_{b,\,n}^{\zg_1,\ldots\zg_p}=0,
\end{align}
for $b=0,\cdots,N-p,\, m=p+b,\cdots, N.$  
Here, by the definition of the tensor $Y_{b,\,n}^{\zg_1,\ldots\zg_p}$,
some special cases have to be distinguished. In particular,
\begin{itemize}
\item {\bf case 1:} \underline{$b$ is even and $n=p$.} 
$Y_{b,\,n}^{\zg_1,\ldots\zg_p}$ is defined by
\begin{align}\label{41.2}
&Y_{b,\,p}^{\zg_1,\ldots\zg_p}=-\varphi_\mu U^\mu 
\frac{4\pi}{(2p+b+1)!}\begin{pmatrix} p+\frac{b}{2} \\p\end{pmatrix}b!p! (2p)!!
\,X_{(0,\,0)}^{\zg_1,\ldots\zg_p}. 
\end{align}
\item {\bf case 2:} \underline{$b$ is even and $n=p+1,\ldots,N.$}
We obtain the following
\begin{align}\label{41.3}
\nonumber &Y_{b,\,n}^{\zg_1,\ldots\zg_p}=-\varphi_\mu U^\mu 
\sum_{r=p+\frac{b}{2}}^{[\frac{(p+b+n)}{2}]}\sum_{s=p}^{r-\frac{b}{2}}\sum_{T=0}^{s-p}
\frac{4\pi}{2r+1}\begin{pmatrix} p+b+n \\2r\end{pmatrix}(-1)^{p+n}
\begin{pmatrix}r\\s\end{pmatrix}\\\nonumber &\cdot \frac{(2r-2s)!}{(2r-2s-b)!}\frac{(n!}{(p+b+n)!}\frac{(2s)!!}{(2s-2p)!!}
\begin{pmatrix}s-p\\T\end{pmatrix}(-1)^T\,X^{\zg_1\ldots \zg_p}_{(n-p-2r+b+2s-2T,\,2r-b-2s+2T)}
\\\nonumber &+\varphi
\sum_{r=p+1+\frac{b}{2}}^{\big[\frac{(p+b+n+1)}{2}\big]}\sum_{s=p}^{r-\frac{b+2}{2}}\sum_{T=0}^{s-p}
\frac{4\pi}{2r+1}\begin{pmatrix} p+b+n+1 \\2r\end{pmatrix}(-1)^{p+n+1}
\begin{pmatrix}r\\s\end{pmatrix}\\ \nonumber &\cdot \frac{(2r-2s)!}{(2r-2s-b-1)!}\frac{n!}{(p+b+n+1)!}\frac{(2s)!!}{(2s-2p)!!}
\\&\begin{pmatrix}s-p\\T\end{pmatrix}(-1)^T\,X^{\zg_1\ldots \zg_p}_{(n-p-2r+b+2s-2T+1,\,2r-b-2s+2T-1)}. 
\end{align}
\item {\bf case 3:} \underline{$b$ is odd and $n=p.$} We have
\begin{align}\label{42.1}
&Y_{b,\,p}^{\zg_1,\ldots\zg_p}=\varphi
\frac{4\pi}{(2p+b+2)!}\begin{pmatrix} p+\frac{b+1}{2} \\p\end{pmatrix}(b+1)!p! (2p)!!
\,X_{(0,\,0)}^{\zg_1,\ldots\zg_p}. 
\end{align}
\item {\bf case 4:} \underline{$b$ is odd and $n=p+1,\ldots,N.$} In this case we get
\begin{align}\label{42.2}
\nonumber &Y_{b,\,n}^{\zg_1,\ldots\zg_p}=-\varphi_\mu U^\mu
\sum_{r=p+\frac{b+1}{2}}^{\big[\frac{(p+b+n)}{2}\big]}\sum_{s=p}^{r-\frac{b+1}{2}}\sum_{T=0}^{s-p}
\frac{4\pi}{2r+1}\begin{pmatrix} p+b+n \\2r\end{pmatrix}(-1)^{p+n+1}
\begin{pmatrix}r\\s\end{pmatrix}\\& \nonumber \cdot \frac{(2r-2s)!}{(2r-2s-b)!}
\frac{n!}{(p+b+n)!}\frac{(2s)!!}{(2s-2p)!!}
\begin{pmatrix}s-p\\T\end{pmatrix}\\\nonumber &
(-1)^T\,X^{\zg_1\ldots \zg_p}_{(n-p-2r+b+2s-2T,\,2r-b-2s+2T)}+
\varphi
\sum_{r=p+\frac{b+1}{2}}^{\big[\frac{(p+b+n+1)}{2}\big]}\sum_{s=p}^{r-\frac{b+1}{2}}
\sum_{T=0}^{s-p}\frac{4\pi}{2r+1}
\\\nonumber&
\cdot \begin{pmatrix} p+b+n+1 \\2r\end{pmatrix}(-1)^{p+n}
\begin{pmatrix}r\\s\end{pmatrix}\frac{(2r-2s)!}{(2r-2s-b-1)!}\frac{n!}{(p+b+n+1)!}
\frac{(2s)!!}{(2s-2p)!!}
\\ &\begin{pmatrix}s-p\\T\end{pmatrix} 
(-1)^T\,X^{\zg_1\ldots \zg_p}_{(n-p-2r+b+2s-2T+1,\,2r-b-2s+2T-1)}
\end{align}
\end{itemize}
Let us begin by considering the system \eqref{41.1} for $b=0,$ so that
it has an equal number
of equations and unknowns. Moreover, the matrix $G_{m,\,n}$ is 
non singular (because in the case
$\varphi=0$ and by the condition $1.$ of hyperbolicity, 
the system must have only 
the zero solution); as a result it yields
\begin{align}\label{43.1} 
Y_{0,\,n}^{\zg_1\ldots\zg_p}=0,\quad \text{for $n=p,\ldots,N$.}
\end{align}
Now we have as an equation \eqref{41.1} with $b=1,\ldots,N-p$ as well as \eqref{43.1}.
In particular, \eqref{43.1} with $n=p,$ for \eqref{41.2} gives
\begin{align}\label{43.2}
\varphi_\mu U^\mu\,X_{(0,\,0)}^{\zg_1,\ldots,\zg_p}=0\,.
\end{align}
From this result we see that $\varphi_\mu U^\mu=0$ is one of the
eigenvalues, because in this case one of the equations is an identity,
so that we have less equations than unknowns to determine the eignvectors.
If we look for other eigenvalues, that is $\varphi_\mu U^\mu\neq 0,$ then 
\eqref{43.2} gives $X_{(0,\,0)}^{\zg_1\ldots\zg_p}=0.$ Using this equation, 
together with \eqref{41.2} and \eqref{42.1}, we get
\begin{align}\label{43.3}
Y_{b,\,p}^{\zg_1,\ldots,\zg_p}=0\,.
\end{align}
so that the term with $n=p$ in the system \eqref{41.1} can be omitted.

Summarizing the results obtained until now, we have found the 
eigenvalue $\varphi_\mu U^\mu=0$ (which does not depend on $G_{m,\,n}$)
and, for the other eigenvalues, the system
\begin{align}\label{44.1}
\begin{cases}
& X_{(0,\,0)}^{\zg_1\ldots\zg_p}=0\,,\\
& Y_{0,\,n}^{\zg_1\ldots\zg_p}=0,\,\, \text{for $n=p,\ldots,N$},\\
&\sum_{n=p+1}^N G_{m,\,n}Y_{0,\,n}^{\zg_1\ldots\zg_p}=0,\,\,\text{for 
$b=1,\ldots N-p$ and $m=p+b,\dots,N$.}
\end{cases}
\end{align}
Let us now repeat the above steps, but with $b=1.$ In this case the 
third equation of \eqref{44.1}
has again an equal number of equations and unknowns 
(because we have dropped $Y_{b,\,p}^{\zg_1\ldots\zg_p}$) and gives the solution 
\begin{align}\label{44.2}
Y_{1,\,n}^{\zg_1\ldots\zg_p}=0,\,\,\,\text{for $n=p+1,\ldots,N$},
\end{align}
and these equations replace the third equation of \eqref{44.1} for $b=1$.
Now we note that by using
\eqref{41.3} and \eqref{42.2}, the second equations of \eqref{44.1} and \eqref{44.2} for $n=p+1$
become
\begin{align*}
\begin{cases}
& \varphi_\mu U^\mu \frac{4\pi}{(2p+1)!}(p+1)!(2p)!!\,X_{(1,\,0)}^{\zg_1\ldots\zg_p}+
\varphi \frac{4\pi}{(2p+3)!}(p+1)(p+1)!(2p)!!2\,X_{(0,\,1)}^{\zg_1\ldots\zg_p}=0,\\
& -\varphi \frac{4\pi}{(2p+3)!}(p+1)2(p+1)!(2p)!!\,X_{(1,\,0)}^{\zg_1\ldots\zg_p}-
\varphi_\mu U^\mu \frac{4\pi}{(2p+3)!}(p+1)2(p+1)!(2p)!!\,X_{(0,\,1)}^{\zg_1\ldots\zg_p}=0.
\end{cases}
\end{align*}
This is a homogeneous system of two equations and two unknowns 
$X_{(1,\,0)}^{\zg_1\ldots\zg_p}$
and $X_{(0,\,1)}^{\zg_1\ldots\zg_p}$, so that we have the following possibilities:
\begin{itemize}
\item If the coefficient matrix is singular, we obtain the 
eigenvalues from the following:
$$(\varphi_\mu U^\mu)^2-\frac{1}{2p+3}\varphi^2=0,$$ which allows us 
to verify that its eigenvalues
do not depend on $G_{m,\,n}$,
\item Its solution is $X_{(1,\,0)}^{\zg_1\ldots\zg_p}=0$
and $X_{(0,\,1)}^{\zg_1\ldots\zg_p}=0$.
\end{itemize}
From this second possibility it follows that
\begin{align}\label{45.1}
Y_{b,\,p+1}^{\zg_1\ldots\zg_p}=0\,,
\end{align}
as seen from \eqref{41.3} and \eqref{42.2} (Note that
$Y_{b,\,n}^{\zg_1\ldots\zg_p}$ is a linear combination of 
$X_{(h,\,k)}^{\zg_1\ldots\zg_p}$ with $h+k=n-p;$ in our case $h+k=1;$
in other words, $Y_{b,\,p+1}^{\zg_1\ldots\zg_p}$ is a linear combination
of $X_{(0,\,1)}^{\zg_1\ldots\zg_p}$ and $X_{(1,\,0)}^{\zg_1\ldots\zg_p}$
which are zero in the present case). From \eqref{45.1} it follows that in 
the third equation of \eqref{44.1} we can omit the term with $n=p+1.$
Summarizing the results of this new step, we have found the set $S^2$ of 
eigenvalues  (the solutions of $\varphi_\mu U^\mu=0$ and of
$(\varphi_\mu U^\mu)^2-\frac{1}{2p+3}\varphi^2=0$) and, to determine
the other eigenvalues, the system
\begin{align}\label{45.2}
\begin{cases}
& X_{(h,\,k)}^{\zg_1\ldots\zg_p}=0\,,\text{for $h+k\leq 1$}\\
& Y_{1,\,n}^{\zg_1\ldots\zg_p}=0,\,\, \text{for $n=p+1,\ldots,N$},\\
&\sum_{n=p+2}^N G_{m,\,n}Y_{b,\,n}^{\zg_1\ldots\zg_p}=0,\,\,\text{for $b=2,
\ldots N-p$ and $m=p+b,\dots,N$.}
\end{cases}
\end{align}
We also note that the set $S^2$ does not depend on $G_{m,\,n}.$ Let us now iterate 
this procedure $\eta$ times and find 
\begin{itemize}
\item A set $S^\eta$ of eigenvalues not depending on $G_{m,\,n}$,
\item The following system, for the determination of other eventual eigenvalues
\begin{align}\label{46.1}
\begin{cases}
& X_{(h,\,k)}^{\zg_1\ldots\zg_p}=0\,,\text{for $h+k\leq \eta-1$}\\
& Y_{q,\,n}^{\zg_1\ldots\zg_p}=0,\,\, \text{for $q=0,\ldots,\eta-1$ and $n=p+q,\ldots,N$},\\
&\sum_{n=p+\eta}^N G_{m,\,n}Y_{b,\,n}^{\zg_1\ldots\zg_p}=0,\,\,\text{for $b=\eta,
\ldots N-p$ and $m=p+b,\dots,N$.}
\end{cases}
\end{align}
\end{itemize}
It remains to prove that, starting from this hypothesis, it follows that it holds
also with $\eta+1$ instead of $\eta.$ The system given by the third equation of \eqref{46.1}
with $b=\eta$ has the solutions 
\begin{align}\label{46.2}
Y_{\eta,\,n}^{\zg_1\ldots\zg_p}=0,\,\,\, \text{for $n=p+\eta,\ldots,N,$}
\end{align}
and these equations replace the third equation of \eqref{46.1} for $b=\eta.$
We also note that by using \eqref{41.3} and \eqref{42.2},
the second equation of \eqref{46.1} and \eqref{46.2} for $n=p+\eta$
constitute a system of homogeneous equations
in the unknowns $X_{(h,\,k)}^{\zg_1\ldots\zg_p}$ with $h+k=\eta.$ Thus we have an 
equal number ($\eta+1$) of equations and of unknowns and this system
does not depend on $G_{m,\,n}.$
By imposing that its coefficient matrix is singular, we obtain 
some eigenvalues which, together with $S^\eta$, constitute the new set $S^{\eta+1}$.
If we look for other eigenvalues, then this system has only 
$X_{(h,\,k)}^{\zg_1\ldots\zg_p}=0$ with $h+k=\eta$ as a solution. On the other hand,
this solution implies $ Y_{b,\,p+\eta}^{\zg_1\ldots\zg_p}=0,$ as seen from 
\eqref{41.3} and \eqref{42.2}. As a result, we can now omit the term with 
$n=p+\eta$ in the third equation of \eqref{46.1}. In other words, we have found 
the set $S^{\eta+1}$ and the system \eqref{46.1} with $\eta+1$ instead of $\eta$. 
This completes the proof of this property.

We note that our system \eqref{41.1} has been gradually replaced by \eqref{43.1}
with $n=p$ at the first step, by the second equation of \eqref{44.1} and \eqref{44.2}
with $n=p+1$ at the second step, and so on. In other words, our system \eqref{41.1}
can be replaced by 
\begin{align}\label{47.1}
Y_{q,\,p+\eta}^{\zg_1\ldots\zg_p}=0,\,\,\text{for $q=0,\ldots,\eta,$}
\end{align}
and on this system we have to impose first the subsystem with $\eta=0$  
to determine $X_{(0,\,0)}^{\zg_1\ldots\zg_p}$, then the subsystem with 
$\eta=1$ to determine $X_{(h,\,k)}^{\zg_1\ldots\zg_p}$ where $h+k=1$,
and so on for increasing values of $\eta.$ In particular, \eqref{47.1},
for fixed $\eta$, will be a subsystem to determine 
$X_{(h,\,k)}^{\zg_1\ldots\zg_p}$ where $h+k=\eta$.

Thus we have proved that the wave speeds do not depend on $G_{m,\,n}$. This fact allows
us to use \eqref{40.1}, without loss of generality. A natural question is the following:
Will this choice satisfy the above condition on the non singularity of the matrix $G_{m,\,n}$
for $m,\,n=p+\eta,\ldots N$? We prove that this is indeed the case. 
First, we see that \eqref{40.1} with the change of integration variable 
$\rho=\frac{k}{\zg}x,$ becomes
\begin{align}\label{48.1}
G_{m,\,n}=\frac{1}{k^2}e^{-\frac{\zl}{k}}\,\int_0^\infty\, 
e^{-\frac{\rho\zg}{k}}\rho^{m+n+2}\,d\rho=
\frac{1}{k^2}e^{-\frac{\zl}{k}}\left(\frac{k}{\zg}\right)^{m+n+3}
\int_0^\infty\,e^{-x}x^{m+n+2}\,dx.
\end{align}
Integrating by parts the expression $\int_0^\infty\,e^{-x}x^p\,dx\,,$ we get
\begin{align*}
\int_0^\infty\,e^{-x}x^p\,dx=\left|-e^{-x}x^p\right|_0^\infty+
p\int_0^\infty\,e^{-x}x^{p-1}\,dx=p\int_0^\infty\,e^{-x}x^{p-1}\,dx.
\end{align*}
Iterating the integration other $p-1$ times we arrive at
\begin{align*}
\int_0^\infty\,e^{-x}x^p\,dx=p!\int_0^\infty\,e^{-x}\,dx=p!.
\end{align*}
As a result, \eqref{48.1} becomes
\begin{align*}
G_{m,\,n}=\frac{1}{k^2}e^{-\frac{\zl}{k}}\left(\frac{k}{\zg}\right)^{m+n+3}\left(m+n+2\right)!\,.
\end{align*}
It follows that
\begin{subequations}\label{49.1}
\begin{align}
\nonumber &\begin{vmatrix}
G_{p+\eta,\,p+\eta} & G_{p+\eta,\,p+\eta+1} & \cdots & G_{p+\eta,\,N}\\
G_{p+\eta+1,\,p+\eta} & G_{p+\eta+1,\,p+\eta+1} & \cdots & G_{p+\eta+1,\,N}\\
\cdots & \cdots & \cdots & \cdots\\
G_{N,\,p+\eta} & G_{N,\,p+\eta+1} & \cdots & G_{N,\,N}
\end{vmatrix}\\\nonumber&{}\\
\nonumber &=\begin{vmatrix}
\frac{(2p+2\eta+2)!}{k^2}\left(\frac{k}{\zg}\right)^{2p+2\eta+3}e^{-\frac{\zl}{k}} 
& \cdots & \frac{(p+\eta+N+2)!}{k^2}\left(\frac{k}{\zg}\right)^{p+\eta+N+3}e^{-\frac{\zl}{k}}\\
\frac{(2p+2\eta+3)!}{k^2}\left(\frac{k}{\zg}\right)^{2p+2\eta+4}e^{-\frac{\zl}{k}} 
& \cdots & \frac{(p+\eta+N+3)!}{k^2}\left(\frac{k}{\zg}\right)^{p+\eta+N+4}e^{-\frac{\zl}{k}}\\
\cdots & \cdots 
& \cdots\\
\frac{(p+\eta+N+2)!}{k^2}\left(\frac{k}{\zg}\right)^{p+\eta+N+3}e^{-\frac{\zl}{k}} 
& \cdots & \frac{(2N+2)!}{k^2}\left(\frac{k}{\zg}\right)^{2N+3}e^{-\frac{\zl}{k}}
\end{vmatrix}
\end{align}
\begin{align}
\nonumber &=
\left(\frac{e^{-\frac{\zl}{k}}}{k^2}\right)^{N-p-\eta+1}\left(\frac{k}{\zg}\right)^
{\displaystyle\sum_{i=3}^{N+1-p-\eta}(2p+\eta+i)}\\\nonumber &\cdot 
\begin{vmatrix}
(2p+2\eta+2)! 
& (2p+2\eta+3)!\frac{k}{\zg} 
& \cdots & (p+\eta+N+2)!\left(\frac{k}{\zg}\right)^{N-p-\eta}\\
(2p+2\eta+3)! 
& (2p+2\eta+4)!\frac{k}{\zg} 
& \cdots & (p+\eta+N+3)!\left(\frac{k}{\zg}\right)^{N-p-\eta}\\
\cdots & \cdots & \cdots & \cdots\\
(p+\eta+N+2)! 
& (p+2\eta+N+3)!\frac{k}{\zg} 
& \cdots & (2N+2)!\left(\frac{k}{\zg}\right)^{N-p-\eta}
\end{vmatrix}\\\nonumber &=
\left(\frac{e^{-\frac{\zl}{k}}}{k^2}\right)^{N-p-\eta+1}\left(\frac{k}{\zg}\right)^
{\displaystyle\sum_{i=3}^{N+1-p-\eta}(2p+\eta+i)+\displaystyle\sum_{j=1}^{N-p-\eta}j}
\\&\cdot 
\begin{vmatrix}
(2p+2\eta+2)! 
& (2p+2\eta+3)! 
& \cdots & (p+\eta+N+2)!\\
(2p+2\eta+3)! 
& (2p+2\eta+4)! 
& \cdots & (p+\eta+N+3)!\\
\cdots & \cdots & \cdots & \cdots\\
(p+\eta+N+2)! 
& (p+2\eta+N+3)!
& \cdots & (2N+2)!
\end{vmatrix}
\end{align}
\end{subequations}
Let us now consider the determinant
\begin{align*}
D_{a,\,d}=
\begin{vmatrix}
a! 
& (a+1)! 
& \cdots & (a+b)!& \cdots & (a+d)!\\
(a+1)! 
& (a+2)! 
& \cdots & (a+b+1)!& \cdots & (a+d+1)!\\
\cdots & \cdots & \cdots & \cdots\\
(a+c)! 
& (a+c+1)! 
& \cdots & (a+b+c)!& \cdots & (a+d+c)!\\
\cdots & \cdots & \cdots & \cdots\\
(a+d)! 
& (a+d+1)! 
& \cdots & (a+b+d)!& \cdots & (a+2d)!
\end{vmatrix}
\end{align*}
and let us sum to the line beginning with $(a+c)!$ the previous one multiplied
by $-(a+c)$. We obtain
\begin{align*}
D_{a,\,d}=
\begin{vmatrix}
a! 
& (a+1)! 
& \cdots & (a+b)!& \cdots & (a+d)!\\
0 
& (a+1)! 
& \cdots & b(a+b)!& \cdots & d(a+d)!\\
\cdots & \cdots & \cdots & \cdots\\
0
& (a+c)! 
& \cdots & b(a+b+c-1)!& \cdots & d(a+d+c-1)!\\
\cdots & \cdots & \cdots & \cdots\\
0 
& (a+d)! 
& \cdots & b(a+b+d-1)!& \cdots & d(a+2d-1)!
\end{vmatrix}
\end{align*}
where we have taken into account that 
$(a+b+c)!-(a+c)(a+b+c-1)!=(a+b+c-1)!(a+b+c-a-c)=b(a+b+c-1)!\,\,$.
It follows that
\begin{align*}
D_{a,\,d}&=a!\,d!
\begin{vmatrix}
(a+1)! 
& \cdots & (a+b)!& \cdots & (a+d)!\\
\cdots & \cdots & \cdots & \cdots\\
(a+c)! 
& \cdots & (a+b+c-1)!& \cdots & (a+d+c-1)!\\
\cdots & \cdots & \cdots & \cdots\\
(a+d)! 
& \cdots & (a+b+d-1)!& \cdots & (a+2d-1)!
\end{vmatrix}\\&=a!\,d!\,D_{a+1,\,d-1}.
\end{align*}
By iterating the procedure $r$ times, we obtain
$$D_{a,\,d}=a!\,(a+1)!\cdots(a-1+r)!\,d!\,(d-1)!\cdots (d+1-r)!\,D_{a+r,\,d-r}.$$
As a result, for $r=d$ this last expression becomes
\begin{align*}
D_{a,\,d}=a!\,(a+1)!\cdots(a-1+d)!\,d!\,(d-1)!\cdots 2!\,\left|(a+d)!\right|\\=
a!\,(a+1)!\cdots(a+d)!\,d!\,(d-1)!\cdots 2!
\end{align*} 
By applying this result with $a=2p+2\eta+2,\,d=N-p-\eta,$ we see that
the determinant in \eqref{49.1} is equal to 
$$(2p+2\eta+2)!\,(2p+2\eta+3)!\,\cdots(p+\eta+N+2)!\,(N-p-\eta)!\,(N-p-\eta-1)!\cdots 2!>0\,.$$
It also follows that the matrix $G_{m,\,n}$ for $m,\,n=0,\ldots, N$ is positive definite.

\section*{Acknowledgements}

This work is supported by Gruppo Nazionale per la Fisica Matematica (GNFM-INdAM) Italy.

\end{document}